# Experimental Realization of an Achromatic Magnetic Mirror based on Metamaterials

GIAMPAOLO PISANO,[1,*] PETER A.R. ADE,[1] CAROLE TUCKER[1]

[1]*School of Physics and Astronomy, Cardiff University, CF24 3AA Cardiff, UK*
*Corresponding author: giampaolo.pisano@astro.cf.ac.uk*

**Our work relates to the use of metamaterials engineered to realize a meta-surface approaching the exotic properties of an ideal object not observed in nature, a 'magnetic mirror'. Previous realizations were based on resonant structures which implied narrow bandwidths and large losses. The working principle of our device is ideally frequency-independent, it does not involve resonances and it does not rely on a specific technology. The performance of our prototype, working at millimetre wavelengths, has never been achieved before and it is superior to any other device reported in the literature, both in the microwave and optical regions. The device inherently has large bandwidth (144%), low losses (<1 %) and is almost independent of incidence-angle and polarization-state and thus approaches the behaviour of an ideal magnetic mirror. Applications of magnetic mirrors range from low-profile antennas, absorbers to optoelectronic devices. Our device can be realised using different technologies to operate in other spectral regions.**

## 1. Introduction

A *Perfect Electric Conductor* (PEC) surface reflects electromagnetic waves, reversing the phase of the electric field while maintaining the phase of the magnetic field (Fig.1(a)). Thus the reflection coefficient for the electric field component is $\Gamma_E$=-1, i.e. a phase change $\Delta\phi_E=\pi$. A PEC surface is well approximated by a metallic mirror. For a *Perfect Magnetic Conductor* (PMC) surface the reflection of the electric and magnetic field components are inverted. The magnetic field vector is reversed in phase while the electric field component remains unchanged (Fig.1(b)). In this case the electric field component reflection coefficient is $\Gamma_E$=+1, i.e. no phase change, $\Delta\phi_E$=0. A PMC surface does not exist naturally. It is possible to model the electric field reflecting off both PEC and PMC surfaces by using simple Transmission Line (TL) circuit models [1]. In these models a PEC is equivalent to the TL terminating in a short circuit (a zero impedance load, $Z_L$=0 (Fig.1(c)), whilst a PMC is equivalent to the TL terminating in an open circuit (an infinite impedance load, $Z_L$=∞ (Fig.1(d)).

Depending on the waveband range of application there are different terminologies used. At optical wavelengths a *Magnetic Mirror* is an artificial device designed to provide electric field in-phase reflection over a defined frequency band. At microwave frequencies, the same component is called an *Artificial Magnetic Conductor* (AMC) surface or *High Impedance Surface*. Since the demonstrator described here is at millimetre wavelengths, we will use the microwave terminology. We define the operational bandwidth of the AMC as the frequency range, normalized to the central frequency $v_0$, where the reflection coefficient phase is kept within $|\Delta\phi|$<90°: $BW_{\pm90}$=($v_{max}$ -$v_{min}$)/$v_0$. In order to approach the ideal behaviour, AMCs are required to work across very large bandwidths with minimal angular dependence, i.e. they should exhibit the same reflection behaviour over a wide range of frequencies and incidence angles.

AMCs were pioneered at microwave frequencies by Sievenpiper [2]. Since then many other devices based on metamaterials have been developed, mainly consisting of Frequency Selective Surfaces (FSSs). These metal-dielectric structures included square, dog-bone and hexagonal patches, mushroom-like geometries, capacitive loaded loops, fractal, Peano and Hilbert curves [3-13]. These designs operated over bandwidths in the range $BW_{\pm90}\sim$10-70% and some of them in multiple narrow bands [5,6]. Microwave applications of AMCs range from low-profile antennas [8-11] to ultrathin absorbers [12,13].

Magnetic mirrors at optical frequencies have been realised more recently using planar fish-scale metallic nanostructures [14]. Since metals at these frequencies suffer high Ohmic losses, a lot of effort is currently devoted to the development of low loss dielectric metamaterials [15-17]. Other realisations include mushroom-like carbon nanotubes, cross- and cube- shaped dielectric resonators [18-20]. These devices find application in optoelectronic devices such as solar cells and have potential use in molecular spectroscopy [14,21]. As for AMCs, the optical magnetic mirrors working principle is based on resonance effects which implies narrow bandwidths (3-30%) and high losses (up to 50%).

Here we present a novel metamaterial AMC device for use at millimetre and sub-millimetre wavelengths. Unlike the inherently narrow-band high-impedance surfaces, a different approach allows us to achieve extraordinarily large operational bandwidths ($BW_{\pm 90}$>140%) superior to any other previous experimental realization, both at microwave and optical frequencies. The device provides a relatively flat reflection phase ($|\Delta\phi|<45°$ across the band) and shows strong angular stability with similar performances up to large off-axis incidence angles ($\vartheta\sim45°$). In addition, the device is nearly polarization independent, i.e. exhibiting very similar performance for the S and P linear polarization components, respectively orthogonal and parallel to the plane of incidence.

Applications of metamaterial AMCs at millimetre and sub-millimetre wavelengths range from retarders, absorbers to phase switching surfaces. These devices would share the same large bandwidths.

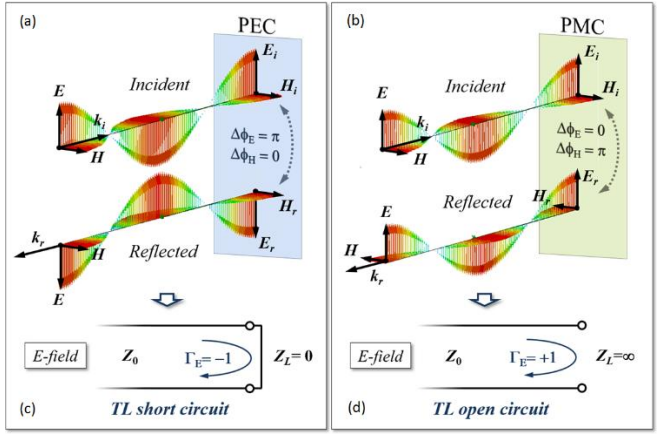

Fig. 1. Reflection of electromagnetic waves on PEC and PMC surfaces. (a) Reflection of an electromagnetic wave on a Perfect Electric Conductor (PEC): E and H field phase-shifts. (b) As above but in the case of a Perfect Magnetic Conductor (PMC). (c) The equivalent transmission line circuit for a PEC surface is a short circuit. It provides out-of-phase reflection coefficient for the electric field. (d) The equivalent transmission line circuit for a PMC surface is a load with infinite impedance. The electric field is reflected without any change in phase.

## 2. AMC development

### A. Conceptual design

The simplest possible AMC design consists of a quarter-wavelength air-gap $\lambda_0/4$, followed by a metallic plane. The half-wavelength extra-path combined with the metallic reflection provides an in-phase reflection at the main frequency, $\nu_0 = c/\lambda_0$, and at its higher harmonics $\nu_n=(2n+1)\nu_0$. The requirement to have thin structures in microwave applications has led to AMC designs based on a metal-dielectric-FSSs. The typical structure is sketched in Fig.2(a) together with its transmission line equivalent circuit. The AMC provides in-phase reflection at a resonance frequency $\nu_0$ and then departs from it more or less rapidly depending on the specific design. The dielectric thickness, starting from the trivial case of $\lambda_0/4$ can be reduced down to $\sim\lambda_0/25$. However, thinner substrates imply narrower bandwidths ($\sim$10% in the latter case).

Although many different sophisticated designs have been proposed, for FSS-based AMCs there are theoretical limits on the maximum achievable bandwidth. This cannot exceed 100%, unless non-negligible losses are allowed within the FSS structure [7]. The AMCs available in the literature have approached but not exceeded this intrinsic limit. In general these devices also systematically show steep in-band phase-shift gradients rapidly departing from the ideal $\Delta\phi=0$. Moreover, these types of AMCs are based on resonances which imply that losses are present.

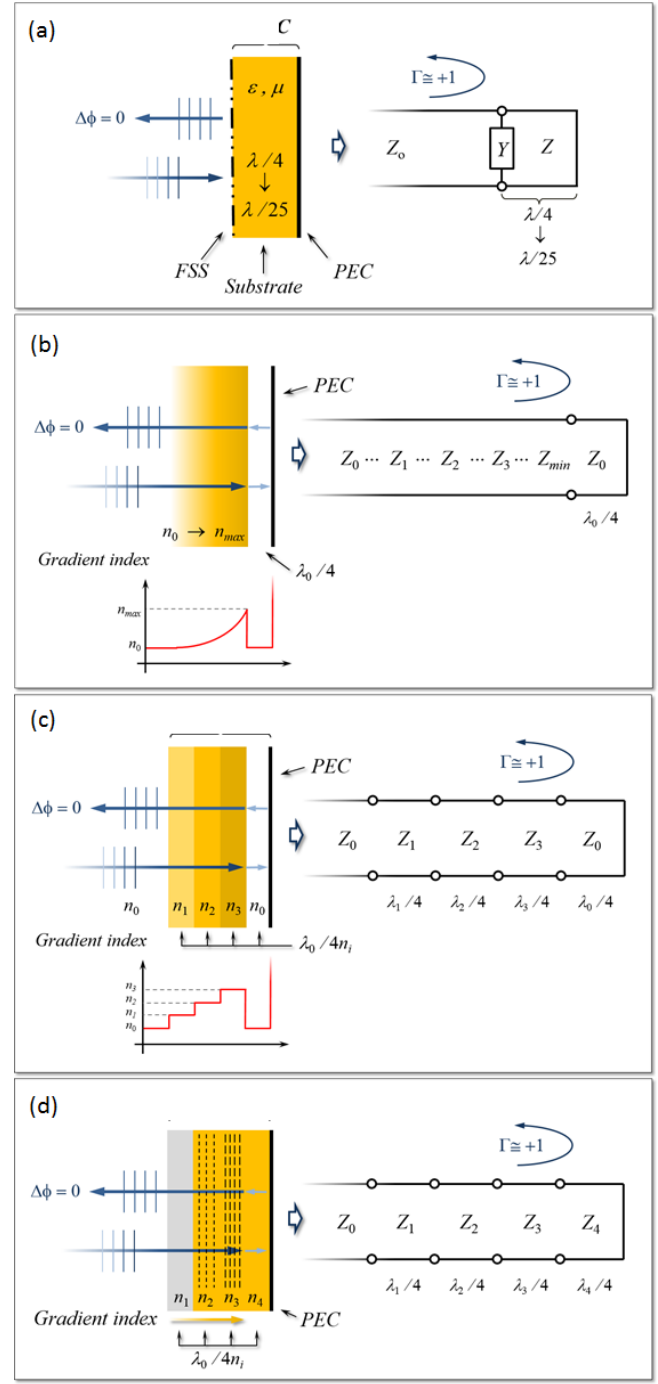

Fig. 2. Classical FSS-based and novel metamaterial-based AMC working principles. (a) Typical FSS based Artificial Magnetic Conductor design and equivalent TL circuit. (b) AMC based on gradient index materials and dielectric internal reflection: the incoming radiation is gently fed into a medium with steadily increasing refractive index until it is almost completely reflected at the interface with free space. In this case the reflection coefficient is $\Gamma$=+1. (c) AMC realized with a discrete number of dielectric layers with quarter-wavelength thicknesses. (d) AMC designed using metamaterials. These layers can be realised by embedding metallic mesh grids inside polymers.

The design proposed in this work is based on a different approach to achieve in-phase reflection. We recall that at the interface between two media with different refractive indices the complex reflection coefficient is given by: $\Gamma = (Z_2 - Z_1)/(Z_1 + Z_2)$, where $Z_{1,2} = Z_0/n_{1,2}$ are the impedances of the first and second medium (assuming no magnetic materials, i.e. $\mu_{r1,2} = 1$) and $Z_0$ is the free space impedance. We notice that depending on the values of $Z_1$ and $Z_2$ the reflection coefficient can be either positive or negative, though not necessarily unitary. Going from low-to-high refractive index ($Z_2<Z_1$) the refection coefficient is negative ($\Gamma < 0$) whereas in the opposite case ($Z_2>Z_1$) the reflection coefficient is positive ($\Gamma > 0$). For the case with radiation travelling into a very much higher refractive index medium ($Z_2<<Z_1$) we will have almost unitary out-of-phase reflection, $\Gamma \approx -1$. In the opposite case, when travelling into a very much lower refractive index ($Z_2>>Z_1$) we obtain almost unitary in-phase reflection $\Gamma \approx +1$. This latter effect is the one adopted for use in our design.

In order to obtain in-phase reflection using this scheme we first need to feed the radiation into the high index medium. This can be obtained by matching the free-space impedance with the higher index material by using a gradient-index medium (Fig.2(b)). After that a sudden jump to a much lower index medium will provide an in-phase reflection $\Gamma \approx +1$. A final metal back-short is added at a distance $\lambda_0/4$ to define the central frequency of operation. The higher the refractive index difference the smaller the transmitted radiation leaking into the gap.

The idealized gradient-index medium can be realized by splitting it into a finite number of discrete quarter-wavelength layers ($\lambda_{n_i}/4$), with specifically optimized refractive indices $n_i$. A sketch of this structure and its TL equivalent circuit are shown in Fig.2(c). In general, the number of available materials suitable at millimetre and sub-millimetre wavelengths is very limited. However, the technology that we have used to produce 'mesh-filters' [22-24] at these frequencies, has also been adopted to realize 'artificial birefringent' waveplates [25,26], and artificial dielectric lenses [27-29]. Here we have extended these developments to realize an AMC device based on metamaterials.

The new AMC design consists of four dielectric layers (Fig.2(d)) that achieve a bandwidth of ~150%. Each layer is a quarter-wavelength thick in its own medium. The first three layers build up the gradient index and the last one replaces the ideal low-index air-gap to avoid implementation difficulties. The structure was initially modelled with a transmission line code that assumed ideal dielectric layers and a central frequency $\nu_0$=240 GHz. The resulting layer refractive indices are: $n_1 \cong 1.2$; $n_2 \cong 1.8$; $n_3 \cong 3.0$; $n_4 \cong 1.5$. Given the material availability at millimetre/sub-millimetre wavelengths, only the first and the last layer can be readily implemented. In our realization the first layer ($n_1$) is made of porous Teflon, a material normally used in anti-reflection coatings, whereas the last one is polypropylene ($n_4$). The middle layers ($n_2$, $n_3$) are artificially built by loading polypropylene with mesh grids. We notice that this configuration dramatically simplifies the manufacture process because three out of the four layers are made by using the same material. We have chosen polypropylene as the basic substrate because of its refractive index, thermal behaviour, availability in multiple thicknesses and the expertise we have built up through its common use in our metal mesh-filter technology [24].

## B. AMC applications

In designing an AMC there is a trade-off between performance parameters that depends upon the type of application. The requirements can be in terms of relative bandwidth, phase flatness, losses and thickness. In our design the in-phase reflection occurs at the end of the gradient index, i.e. at the high-to-low index interface. Keeping this in mind, we can distinguish three types of application:

I) Applications where the in-phase (PMC) reflection is required to happen at a specific plane within a medium (Fig.3(a)-I). For example, a Salisbury screen absorber can be simply implemented loading the high-to-low index interface within the AMC with a resistive layer.

II) Applications where an in-phase (PMC) reflection is used in combination with an out-of-phase (PEC) reflection and both happen within the same medium and at the same reference plane (Fig.3(a)-II). This can be achieved by adding metal patterns at the high-to-low index interface to obtain the PEC reflection where required. Phase switching modulation (0-180°) can be easily implemented in this way.

III) Applications where an in-phase (PMC) reflection is required to happen at a specific reference plane in free space (Fig.3(a)-III). The 'equivalent' free-space reference plane, where the PMC reflection happens, is identified taking into account the real optical path within the gradient index medium. As an example, an interferometer where the radiation in one arm is alternatively reflected by a PEC or PMC surface would subsequently switch the role of the output ports.

In this work, we have developed a device of the type-II above. It includes both embedded PMC and PEC structures with the same reflection plane. This facilitates the differential phase measurement that will just require probing two areas of the surface. The same device, although not optimized for it, can also be used for a type-III application provided the equivalent reference plane is identified.

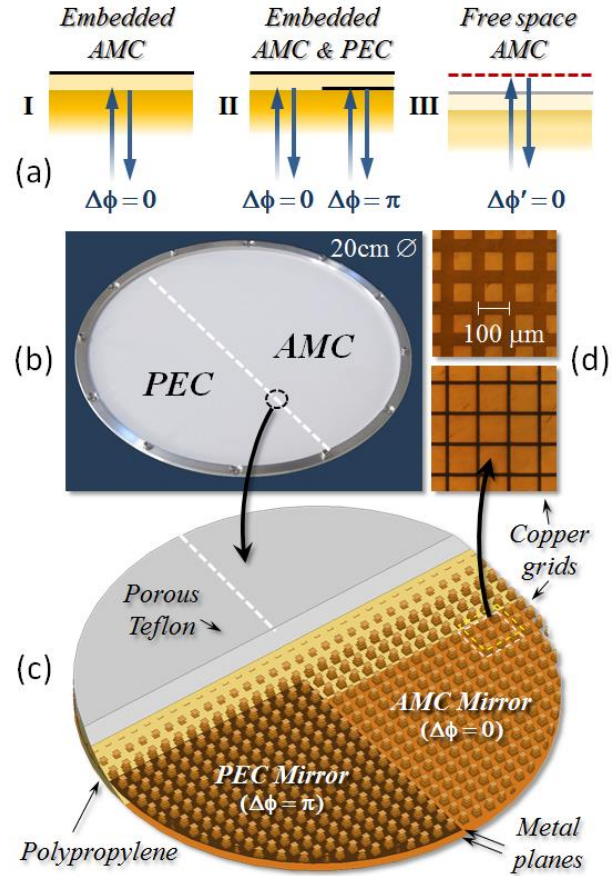

Fig. 3. AMC applications and prototype details. (a) Different types of application for an AMC. (b) Picture of the AMC prototype: half of it behaves like an embedded AMC, whereas the other half is an embedded PEC. The white material is the lowest index material used in the design. (c) Sketch of a portion of the metamaterial structures. (d) Photographs of the two types of embedded metal grids used in the prototype.

## C. AMC modelling and prototype

The device that we have prototyped is shown in Fig.3(b). It is divided in two parts, as sketched and detailed in Fig.3(c). Half of its surface is an embedded AMC, whereas the other half is an embedded PEC. Finite element analysis (FEA) was required to accurately model these metamaterial structures [30], since they are based on polypropylene embedded copper grids (Fig.3(d)).

The results of the on-axis finite-element simulations are shown in Fig.4. The embedded AMC phase-shift is kept within ±90° across an extremely large bandwidth, of the order of $BW_{\pm 90}\sim 147\%$. In FSS-based AMCs the phase crosses zero at the central frequency and moves away from it almost linearly (typical curve and theoretical limit in Fig.4). In our device, the phase oscillates around zero across a large frequency range to then reach ±90° with losses <1%. The new device can also be used in a free space application (type-III).

The performance of the AMC design discussed so far was modelled assuming radiation at normal incidence, a condition that might not be always satisfied. In addition, real optical beams have a finite divergence, spreading the angles of incidence around the off-axis value. For these reasons it is important for the AMC not only to be able to work off-axis but also to have stable performance across a wide range of incidence angles. We have therefore investigated the off-axis performance of the new AMC using the FEA models. The results show the phase properties are maintained with strong angular stability for incidence angles up to 45 degrees (as shown later in Fig.6(b) and Fig.6(c)).

Using the same FEA models it was possible to calculate the absorption coefficients of the embedded AMC and PEC surfaces. Within the $BW_{\pm 90}$ frequency range the resulting average absorption coefficients were very small: $\alpha_{AMC}=0.007$ and $\alpha_{PEC}=0.010$. The AMC absorption is slightly lower than the PEC one because the radiation is mostly reflected at the high-to-low dielectric interface and barely interacts with the finite conductivity back-short mirror.

The embedded AMC prototype was manufactured using photolithographic techniques that are well established for multilayer mesh-filters production as used in many astronomical instruments from far-infrared to millimetre wavelengths [24]. The specific technique adopted for the AMC consisted of embedding the metal grids within polypropylene. This can be achieved by hot-pressing a stack of ordered grids and polypropylene layers in order to obtain a final unique robust block of material with the metallic patches suspended within it.

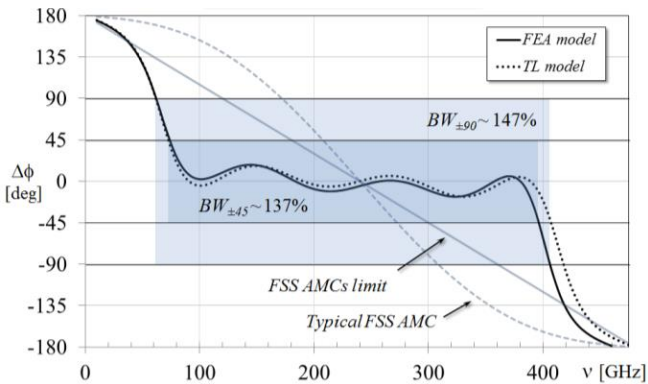

Fig. 4. AMC modelled performances. Embedded AMC on-axis phase-shift simulations compared to the typical performance and to the theoretical limit of FSS based AMCs.

## 3. AMC experimental characterization

### A. Experimental setup

The device had a diameter of 20 cm and it was 745 μm thick. It was characterized at room temperature using two independent measurement techniques based on coherent and incoherent sources: a Vector Network Analyser (VNA) and a Fourier Transform Spectrometer (FTS). These were used in the two different optical configurations, as shown in Fig.5. The coherent tests were carried out using a *Rohde-Schwarz ZVA67* VNA equipped with heads working in the 75-110 GHz and 170-260 GHz ranges. The incoherent tests were performed by using a Fourier Transform Spectrometer which could operate continuously over a wide range of frequencies: 130-500 GHz. The AMC was tested off-axis, at ϑ=45° incidence angle, within the optical setup sketched in Fig.5(a). The AMC was mounted on a rotation stage allowing differential measurements of the PEC and PMC surfaces via a 180° rotation around its axis. The AMC phase-shift was calculated in the same way as discussed for the modelling: a value of π was subtracted from the differential phase-shift between the PEC and the AMC surfaces.

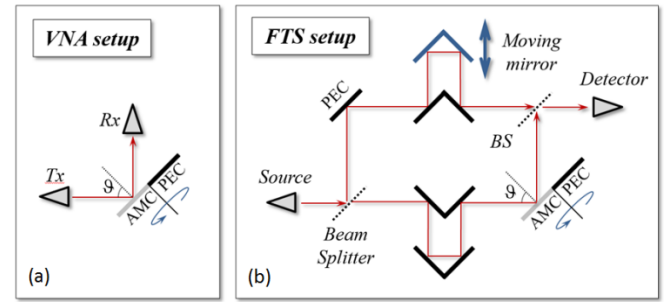

Fig. 5. AMC optical test setups. In both the VNA (a) and FTS (b) setups the prototype was testes in reflection at 45 degrees incidence angle. The device was able to rotate around its optical axis in order to operate in the AMC (null phase-shift) or PEC (π phase-shift) modes. In the VNA setup the change in phase was acquired directly by the receiver head. In the modified Mach-Zehnder FTS setup the device replaces one of the optical mirrors and the change in phase results in the inversion of the inteferogram acquired by the detector.

### B. AMC measurements and results

The FTS data were acquired by using a cryogenically cooled bolometric detector working at 1.6K. However, the FTS signal-to-noise ratio compared to that of the VNA is low because it uses a thermal Mercury arc lamp source which decreases in intensity as $\nu^2$ and because the throughput of the modified Mach-Zehnder interferometer is limited.

The FTS detected signal is an intensity interferogram, i.e. the signal resulting from the interference of two beams propagating across the two arms of the interferometer, one of which has its optical path length varied during the acquisition. In normal FTS configurations, the device under test is positioned after the recombination of the two signals. Interferograms are acquired with and without the device and then Fourier transformed to get the relative spectra. The final spectrum is obtained by normalizing the device spectrum with the background one. This configuration provides either the intensity transmission or reflection as a function of frequency.

In our case, in order to get the phase information it was necessary to modify the classical FTS optical scheme and allow the device under test to be placed within one arm of the interferometer. In this case the reflection phase could be directly measured by positioning the device in place of one of the mirrors in one arm of a Mach-Zehnder interferometer configuration as shown in Fig.5(b). By rotating our device by 180 degrees around its optical axis it was possible to alternatively reflect the beam off the PEC or PMC surfaces. With the PEC side reflecting the beam, the usual π phase-shift is effected and the FTS will work in the usual way such that to all the frequency components within its operational bandwidth are constructively in-phase when the arm path lengths are the same. By rotating the device by 180 degrees, the AMC surface will now reflect the beam providing a null phase-shift to all the frequency components over its band, with the net effect that the interferogram is inverted (switched in phase) as each Fourier component now interferes destructively. However, the AMC does not provide exactly null phase-shifts and the interferogram will not necessarily be an exact mirror image of the previous PEC reflector case (see Fig.6(a)). In order to extract the phase information it is necessary to compute the complex Fourier transform of the two interferograms, calculate their complex arguments and then subtract them. The differential phase-shift between the PEC and AMC surface will be:

$$\text{Arg(FFT}_{PEC}) - \text{Arg(FFT}_{AMC}) \tag{1}$$

The AMC phase-shift $\Delta\phi$ is obtained by subtracting a value of π. Note that the FTS data spans almost the whole AMC operational bandwidth.

The AMC phase-shifts measured with the VNA and with the modified Mach-Zehnder FTS are shown in Fig.6(b) and Fig.6(c) respectively for the *S* and *P* polarizations. In both types of measurements the AMC was tested at 45° incidence. The good agreement between the model and the experimental measurements proves the working principle and the performance of our device. The phase-shift is kept within ±90° across the ~75-460 GHz frequency range for both the *S* and *P* polarisations. This is equivalent to a relative bandwidth $BW_{\pm90(S,P)}$~144%, never achieved before and well above the 100% theoretical limit of the FSS-based AMCs. In addition, within the band the phase oscillates several times around the zero value, not crossing it just once as in the FSS cases.

We note that there is a small systematic shift between the model and the FTS/VNA data, in both S and P polarizations. On the one hand, some uncertainties in the model input parameters might contribute to this. On the other hand, the critical alignment between the device and rotary stage axes would also account for some of the discrepancy.

## 4. Conclusions

We have developed a new type of 'Artificial Magnetic Conductor' surface or 'Magnetic Mirror', so called respectively within the microwave and the optical terminologies. Our device, designed to work at millimetre wavelengths, can provide the required 'in-phase' reflection over extremely large bandwidths, up to large incidence angles and with very low associated losses.

In microwave engineering, AMCs are mainly based on Frequency Selective Surfaces (FSSs), whereas in the optical region the development of magnetic mirrors, started with lossy metallic structures, is now migrating into dielectric metamaterials with lower losses. In both cases, the magnetic mirrors are based on resonant structures which imply efficient operation only within narrow bandwidths and associated large losses.

The working principle of our new device is completely different from any previous realization being based on internal reflections in high permittivity dielectric materials. In this case the 'in-phase' reflection is ideally frequency-independent and there are no resonances involved in the process. This working principle is 'technology-independent' and can be used to design AMCs/magnetic-mirrors (or other applications employing them) working in other spectral regions and realized with different technologies.

In our specific design, both the required high permittivity materials and the gradient index medium are synthesized by using metamaterials.

We have used the mesh-technology because it is the one we are familiar with in developing other quasi-optical devices, in the mm and sub-mm range. Our grids do not work in the lossy and bandwidth-limited regime of the FSSs.

The performance of the presented prototype have never been achieved before. Its 144% bandwidth is vastly superior to any other reported device available in the literature. It is superior to the theoretical limit of FSS-based AMCs [7] and to any other magnetic mirrors realized so far. In addition, within the operational bandwidth, the phase is relatively flat, compared to the usual monotonic behaviour of other existing devices. Our device is almost polarisation independent and works up to more than 45 degree angles of incidence. All the above properties (bandwidth, low losses, angle and polarisation independence) allow our AMC surface to approach the ideal behaviour of a Perfect Magnetic Conductor surface.

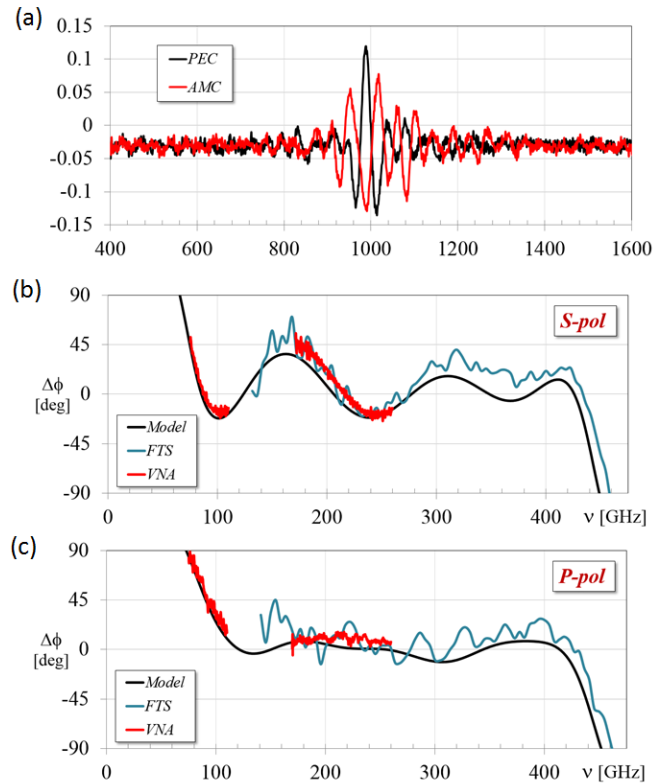

Fig. 6. AMC experimental results. (a) FTS interferograms for the PEC (black) and AMC (red) parts of the device (arbitrary units on both axes). Notice how the AMC interferogram looks out-of-phase compared to the PEC one. (b) and (c) show the VNA and FTS phase-shifts measurements against model expectations respectively for the S and P polarisation. The AMC operated at 45 degrees incidence angle.

**Funding Information.** Science and Technology Facility Council (STFC) Cardiff Consolidated Grants (ST/M007707/1 and ST/K00926/1).

**Acknowledgment**. The Authors acknowledge support from the Science and Technology Facility Council through the current Consolidated Grants.

## References

1. D.M. Pozar, "Microwave Engineering," (John Wiley & Sons, 2011)
2. D. Sievenpiper, Z. Lijun, R.F.J. Broas, N.G. Alexopolous, E. Yablonovitch, "High-Impedance Electromagnetic Surfaces with a Forbidden Frequency Band," IEEE Trans. Micr. Theo. Tech. **47**, 2059 – 2074 (1999).
3. S. Clavijo, R.E. Diaz, W.E.M. McKinzie, "Design Methodology for Sievenpiper High-Impedance Surfaces: An Artificial Magnetic Conductor for Positive Gain Electrically Small Antennas," IEEE Trans. Ant. Prop. **51**, 2678–2690 (2003).
4. A. Erentok, P. L. Luljak, and R. W. Ziolkowski, "Characterization of a Volumetric Metamaterial Realization of an Artificial Magnetic Conductor for Antenna Applications," IEEE Trans. Ant. Prop. **53** , 160 - 172 (2005).
5. D. J. Kern, D. H. Werner, A. Monorchio, L. Lanuzza, and M. J.Wilhelm, "The Design Synthesis of Multiband Artificial Magnetic Conductors Using High Impedance Frequency Selective Surfaces," IEEE Trans. Ant. Prop. **53** , 8 – 17 (2005).
6. M.E. de Cos, Y. Álvarez, F. Las-Heras, "Novel Broadband Artificial Magnetic Conductor With Hexagonal Unit Cell," IEEE Antennas and Wireless Propagation Letters **10**, 615-618 (2011).
7. C.R. Brewitt-Taylor, "Limitation on the bandwidth of artificial perfect magnetic conductor surfaces," IET Micr. Ant. Prop. **1**, 255–260 (2007).
8. A. P. Feresidis, G. Goussetis, S. H. Wang, and J. C. Vardaxoglou, "Artificial Magnetic Conductor Surfaces and Their Application to Low-Profile High-Gain Planar Antennas," IEEE Trans. Ant. Prop. **53**, 209 - 215 (2005).
9. J. R. Sohn, K. Y. Kim, H.-S. Tae, J.-H. Lee, "Comparative Study on Various Artificial Magnetic Conductors for Low-Profile Antenna," PIER **61**, 27–37 (2006).
10. A. Vallecchi, J. R. De Luis, F. Capolino, and F. De Flaviis, "Low Profile Fully Planar Folded Dipole Antenna on a High Impedance Surface," IEEE Trans. Antennas Propag. **60**, 51 - 62 (2012).
11. K. Agarwal, Nasimuddin, A. Alphones, "Wideband Circularly Polarized AMC Reflector Backed Aperture Antenna," IEEE Trans. Ant. Prop. **61**, 1456-1461 (2013).
12. D.J. Kern, D.H. Werner, "Magnetic Loading of EBG AMC Ground Planes and Ultrathin Absorbers for Improved Bandwidth Performance and Reduced Size," Micr. Opt. Tech. Lett. **48**, 2468-2471 (2006).
13. F. Costa, A. Monorchio, G. Manara, "Analysis and Design of Ultra Thin Electromagnetic Absorbers Comprising Resistively Loaded High Impedance Surfaces," IEEE Trans. Ant. Prop., **58**, 1551-1558 (2010).
14. A.S. Schwanecke V.A. Fedotov, V.V. Khardikov, S.L. Prosvirnin, Y. Chen and N.I. Zheludev, "Optical magnetic mirrors," J. Opt. A **9**, L1-L2 (2007).
15. J. C. Ginn et al., "Realizing Optical Magnetism from Dielectric Metamaterials," Phys. Rev. Lett. **108** , 097402 (2012).
16. L. Shi, T. U. Tuzer, R. Fenollosa, F. Meseguer, "A New Dielectric Metamaterial Building Block with a Strong Magnetic Response in the Sub-1.5-Micrometer Region: Silicon Colloid Nanocavities," Adv. Mater. **24** , 5934 – 5938 (2012).
17. A. I. Kuznetsov et al., "Magnetic light," Sci. Rep. **2** , 492 (2012).
18. H. Rostami, Y. Abdi, and E. Arzi, "Fabrication of optical magnetic mirrors using bent and mushroom-like carbon nanotubes," Carbon **48** , 3659 - 3666 (2010).
19. J. Z. Hao, Y. Seokho, L. Lan, D. Brocker, D. H. Werner, T. S. Mayer, "Experimental Demonstration of an Optical Artificial Perfect Magnetic Mirror Using Dielectric Resonators," IEEE Antennas and Propagation Society International Symposium, 1 - 2 (2012).
20. S. Liu et al., "Optical magnetic mirrors without metals," Optica **1**, 250-256 (2014).
21. M. Esfandyarpour, "Metamaterial mirrors in optoelectronic devices," Nature Nanotechnology **9**, 542-547 (2014).
22. N. Marcuvitz, "Waveguide Handbook," M.I.T. Rad. Lab. Ser., Mc.Graw-Hill (1951), 280-290.
23. R. Ulrich , "Far-infrared properties of metallic mesh and its complementary structure" Infrared Physics, **7**,1 pp.37-50.
24. P. A. R. Ade, G. Pisano, C. E. Tucker, S. O. Weaver, "A Review of Metal Mesh Filters," Proceedings of the SPIE, **6275**, pp.U2750 (2006).
25. G. Pisano, G. Savini, P.A.R. Ade, V. Haynes, "A Metal-mesh Achromatic Half-Wave Plate for use at Submillimetre Wavelengths," Appl. Opt. **47**, 6251-6256 (2008).
26. G.Pisano, M. W. Ng, V. Haynes and B. Maffei, "A Broadband Metal-Mesh Half-Wave Plate for Millimetre Wave Linear Polarisation Rotation," Progress In Electromagnetics Research M, **25**, pp.101-114 (2012)
27. G. Pisano, M.W. Ng, B. Maffei, F. Ozturk, "A Dielectrically Embedded Flat Mesh Lens for Millimetre Waves Applications," Appl. Opt. **52**, 2218-2225 (2013).
28. Zhang, J., Ade, P.A.R, Mauskopf, P., Moncelsi, L., Savini, G. and Whitehouse, N., "New Artificial Dielectric Metamaterial and its Application as a THz Anti-Reflection Coating," Appl. Opt. **48**, 6635-6642 (2009).
29. G. Savini, P.A.R. Ade, J. Zhang, "A New Artificial Material Approach for Flat THz Frequency Lenses," Opt. Expr. **20**, 25766-25773 (2012).
30. High Frequency Structure simulator (HFSS): www.ansys.com